\documentclass[aps,prl,reprint,groupedaddress,floatfix]{revtex4-1}

\usepackage{graphicx}
\def\missET {{\not\!\! E_T}}

\begin{document}

\title[ Enhanced Higgs to $\tau^+\tau^-$ Searches with Deep Learning ]{Enhanced Higgs to $\tau^+\tau^-$ Search with Deep Learning }

\author{ P. Baldi}
\address{Dept. of Computer Science, UC Irvine, Irvine, CA 92617}
\author{ P. Sadowski}
\address{Dept. of Computer Science, UC Irvine, Irvine, CA 92617}
\author{ D. Whiteson}
\address{Dept. of Physics and Astronomy, UC Irvine, Irvine, CA 92617}
\begin{abstract}
The Higgs boson is thought to provide the interaction that imparts mass to the fundamental fermions, but while measurements at the Large Hadron Collider (LHC) are consistent with this hypothesis, current analysis techniques lack the statistical power to cross the traditional 5$\sigma$ significance barrier without more data. \emph{Deep learning} techniques have the potential to increase the statistical power of this analysis by \emph{automatically} learning complex, high-level data representations. In this work, deep neural networks are used to detect the decay of the Higgs to a pair of tau leptons. A Bayesian optimization algorithm is used to tune the network architecture and training algorithm hyperparameters, resulting in a deep network of eight non-linear processing layers that improves upon the performance of shallow classifiers even without the use of features specifically engineered by physicists for this application. The improvement in discovery significance is equivalent to an increase in the accumulated dataset of 25\%.
\end{abstract}

\maketitle

\section{Introduction}

Observations made at the LHC led to the announcement of the discovery of the Higgs boson in 2012~\cite{aad_particle_2012,abbaneo_new_2012}, and much more data will be collected when the collider comes back online in 2015. A top priority is to demonstrate that the Higgs boson couples to fermions through direct decay modes. Of the available modes, the most promising is the decay to a pair of tau leptons ($\tau^\pm$), which balances a modest branching ratio with manageable backgrounds.  From the measurements collected in 2011-2012, the LHC collaborations report data consistent with $H\rightarrow\tau^+\tau^-$ decays, but without statistical power to cross the $5\sigma$ threshold, the standard for claims of discovery in high-energy physics. There is a vigorous effort in the high-energy physics community to improve the statistical analysis of collider data in order to require smaller accumumated datasets for this scientific discovery and others like it.

Machine learning is already widely used in the $H\rightarrow\tau^+\tau^-$ search and other areas of high-energy physics, but standard software packages primarily rely on shallow learning models such as artificial neural networks with only a single hidden layer. Recent interest in \emph{deep learning} has resulted in significant advances in computer vision and speech recognition. This technique uses deep neural networks with multiple non-linear hidden layers, which are able to represent complex functions more efficiently than shallow networks, and may generalize better to new data due to architectural constraints~\cite{montufar_number_2014}. While training such networks is notoriously difficult due to the vanishing gradient problem~\cite{hochreiter_recurrent_1998,bengio1994learning}, recent advances in computing hardware have made it feasible to train deeper networks on larger datasets. 

In the field of high-energy physics, deep neural networks have demonstrated an ability to significantly increase classification performance and discovery significance on two other high-energy physics applications~\cite{baldi_searching_2014}.  This is important, as the colliders are expensive to operate and the particle detection systems have short lifetimes due to the intense radiation produced in collisions. Therefore, boosting the discovery significance can shorten the time needed to make a discovery, or make discoveries possible in limited-size datasets.  Because classifiers can be trained on an arbitrary quantity of simulated data, large neural network architectures with millions of parameters can be trained without overfitting. Thus, the challenge lies in selecting an appropriate set of hyperparameters that determine the network architecture and training algorithm details that yield the best-fitting classifier.

In this Letter, we apply deep learning techniques to the important application of detecting $H\rightarrow\tau^+\tau^-$ decays, where it is worth the computational cost to tune these parameters carefully in order to maximize the statistical power of the analysis. The hyperparameters are optimized systematically using a Bayesian optimization algorithm.

\begin{figure}
\centering
\includegraphics[width=2.5in]{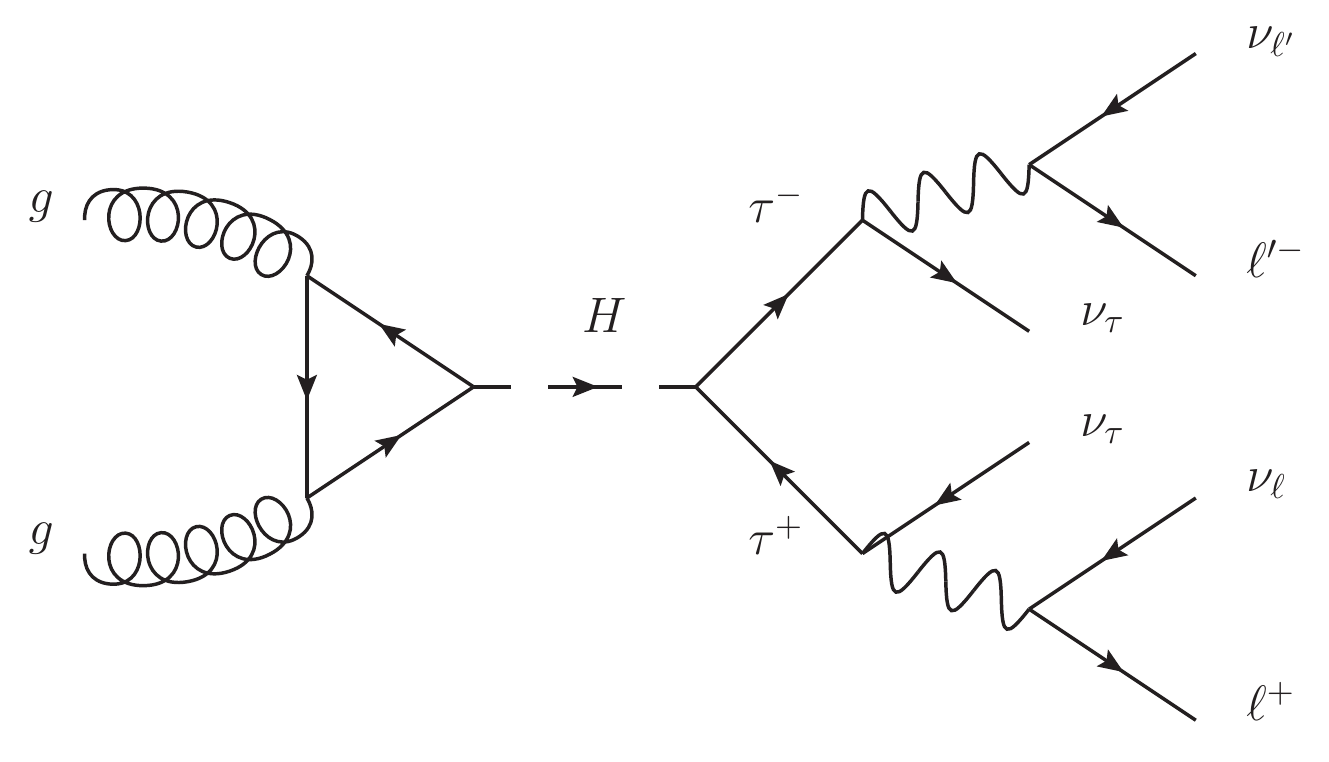}\\
\includegraphics[width=2.5in]{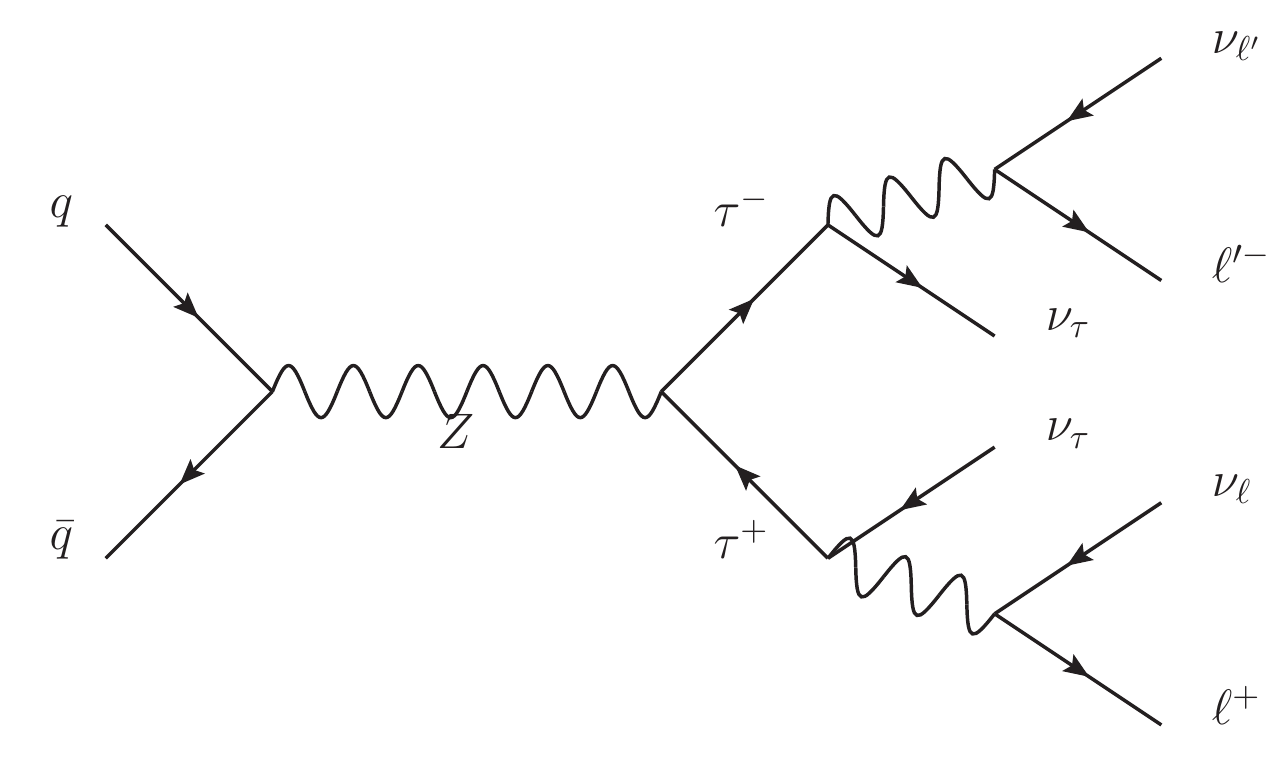}
\caption{ Feynman diagrams describing the signal $gg\rightarrow H\rightarrow\tau^+\tau^-\rightarrow \ell^-\nu\nu\ell^+\nu\nu$ process and the dominant background $q\bar{q}\rightarrow Z\rightarrow\tau^+\tau^-\rightarrow \ell^-\nu\nu\ell^+\nu\nu$ process.}
\label{fig:diag}
\end{figure}

\section{Model}

Proton collisions at the LHC annhiliate the proton constituents, quarks, and gluons. In a small fraction of these collisions, a new heavy state of matter forms, such as a Higgs or $Z$ boson. These heavy states are unstable, quickly decaying into successively lighter and more stable particles. In the case of a Higgs boson, the decay process is:

\begin{equation}
gg\rightarrow H \rightarrow \tau^+\tau^-
\end{equation}

In the study presented here, the decay of $\tau$ leptons into lighter leptons ($e$ and $\mu$) and pairs of neutrinos ($\nu$), $\tau^\pm\rightarrow\ell^\pm\nu_\tau\nu_\ell$ is considered; see Fig.~\ref{fig:diag}. This is the most challenging decay mode, as it involves the largest number of invisible neutrinos.

Detectors surrounding the point of collision measure the identity, momentum, and direction of the visible final stable particles; the intermediate states of matter are not observable. Two processes that generate the same sets of stable particles can be difficult to distinguish. Figure~\ref{fig:diag} shows how the process $q\bar{q}\rightarrow Z\rightarrow \tau^+\tau^-$ yields the identical list of particles as a process that produces the Higgs boson.

To distinguish between two processes with identical final state particles, the momentum and direction of the visible final state particles are examined closely. Given perfect measurement resolution and a complete description of the final state particles $B$ and $C$, the invariant mass of the short-lived intermediate state $A$ in the process $A\rightarrow B+C$ is given by:

\begin{equation}
 m^2_{A} = m^2_{B+C} = (E_B + E_C)^2 - |(p_B+p_C)|^2 
\end{equation}
 
However, finite measurement resolution and escaping neutrinos (which are invisible to the detectors) make it impossible to calculate the intermediate state mass precisely. Instead, the momentum and direction of the final state particles are studied. Sophisticated Monte Carlo programs have been carefully tuned to produce highly faithful collision simulations, and simulated data is used to investigate methods for distinguishing between possible generating processes. In the studies shown here, all samples are generated with the {\sc madgraph}5~\cite{madgraph} program, with showering and hadronization performed by {\sc pythia}~\cite{pythia} and ATLAS detector simulation with {\sc delphes}~\cite{delphes}.  Machine learning can then be applied to learn a classification model from this simulated data. These classifiers take as input the variables that are measured by the detectors (and/or high-level variables that are derived from these measurements), and learn to predict the probability that a given example was the result of a particular generating process. The relevant variables for $H\rightarrow\tau^+\tau^-$ classification are described below.

\subsection{Low-level variables}

Ten essential measurements are provided by the detectors:

\begin{itemize}
\item The three-dimensional momenta, $p$, of the charged leptons;
\item The imbalance of transverse momentum ($\missET$) in the final state transverse to the beam direction, due to unobserved or mismeasured particles;
\item The number and momenta of particle `jets' due to radiation of gluons or quarks.
\end{itemize}

\begin{figure}
\centering
\includegraphics[width=1.5in]{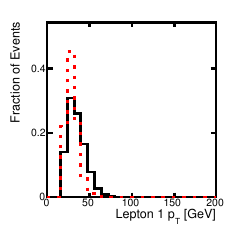}
\includegraphics[width=1.5in]{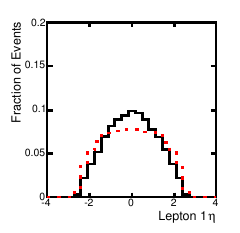}
\includegraphics[width=1.5in]{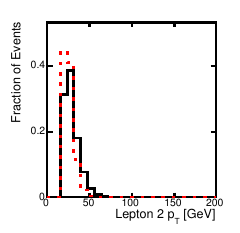}
\includegraphics[width=1.5in]{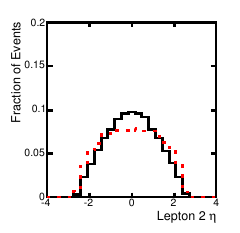}
\includegraphics[width=1.5in]{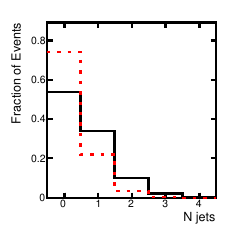}
\includegraphics[width=1.5in]{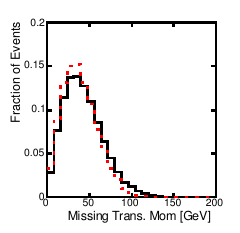}
\caption{ Distributions of low-level input variables from basic kinematic quantities in $\ell\ell+\missET$ events for simulated signal (black, solid) and background (red, dashed) benchmark events. Shown are the distributions of transverse momenta ($p_{\rm T}$) and azimuthal angle ($\eta$)  of each observed particle as well the number of hadronic jets ($N_{\rm jets}$) and the imbalance of transverse momentum ($\missET$) in the final state. Polar angle ($\phi$) information for each observed particle is also available to the network, but is not shown, as the one-dimensional projections have little information. }
\label{fig:llvar}
\end{figure}

Distributions of these variables in simulation are given in Fig.~\ref{fig:llvar}.

\subsection{High-level variables}

In order to better discriminate between Higgs-boson production and $Z$-boson production, there is a vigorous effort to construct non-linear combinations of these low-level variables that capture useful high-level information. The derived variables that have been considered include:

\begin{itemize}
\item Axial missing momentum, $\missET\cdot p_{\ell^+\ell^-}$;
\item Scalar sum of the observed momenta, $|p_{\ell^+}|+|p_{\ell^-}|+|\missET|+\sum_i |p_{\textrm{jet}_i}|$;
\item Relative missing momentum, $\missET$ if $\Delta\phi(p,\missET)\ge\pi/2$, and $\missET\times\sin(\Delta\phi(p,\missET)$ if $\Delta\phi(p,\missET)<\pi/2$, where $p$ is the momentum of any charged lepton or jet;
\item Difference in lepton azimuthal angles, $\Delta\phi(\ell^+,\ell^-)$;
\item Difference in lepton polar angles, $\Delta\eta(\ell^+,\ell^-)$;
\item Angular distance between leptons, $\Delta R=\sqrt{(\Delta\eta)^2+(\Delta\phi)^2}$;
\item Invariant mass of the two leptons, $m_{\ell^+\ell^-}$;
\item Missing mass, $m_{\textrm{MMC}}$~\cite{Elagin:2010aw};
\item Sphericity and transverse sphericity;
\item Invariant mass of all visible objects (leptons and jets).
\end{itemize}

\begin{figure}
\centering
\includegraphics[width=1.25in]{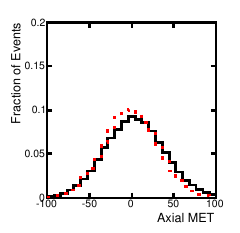}
\includegraphics[width=1.25in]{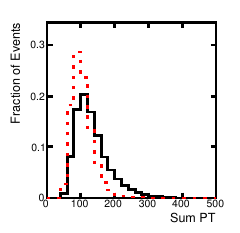}
\includegraphics[width=1.25in]{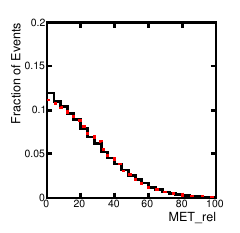}
\includegraphics[width=1.25in]{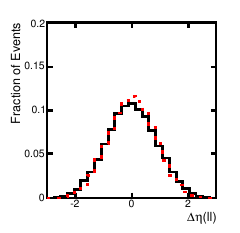}
\includegraphics[width=1.25in]{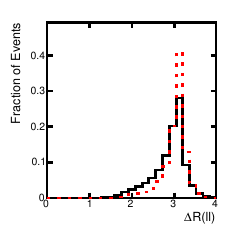}
\includegraphics[width=1.25in]{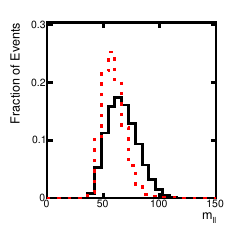}
\includegraphics[width=1.25in]{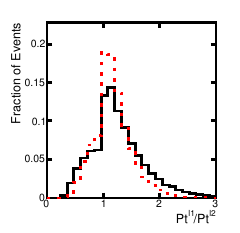}
\includegraphics[width=1.25in]{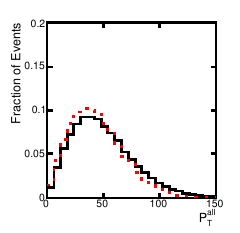}
\includegraphics[width=1.25in]{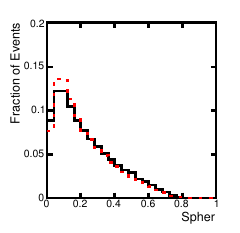}
\includegraphics[width=1.25in]{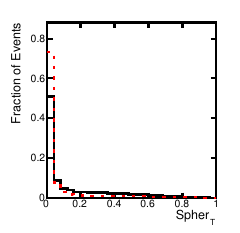}
\includegraphics[width=1.25in]{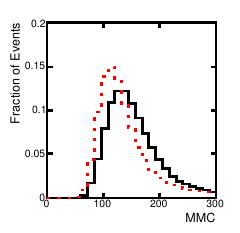}
\includegraphics[width=1.25in]{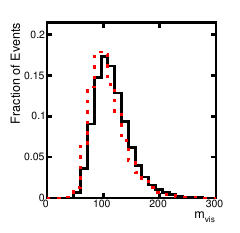}
\includegraphics[width=1.25in]{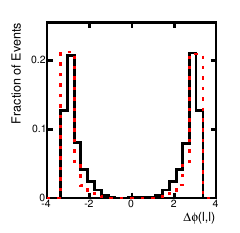}
\caption{  Distributions of high-level input variables derived from the low-level variables in $\ell\ell+\missET$ events for simulated signal (black, solid) and background (red, dashed) benchmark events. See text for definitions. }
\label{fig:hlvar}
\end{figure}

Distributions of these variables in simulation are given in Fig.~\ref{fig:hlvar}.  At first glance, these high-level variables appear to contain more discriminatory power than the low-level variables. 

\section{Results}

Deep neural networks were trained with state of the art techniques to detect $H\rightarrow\tau^+\tau^-$ decay events using a Bayesian optimization algorithm to select hyperparameters. The results are compared to a similar optimization over shallow neural networks with the same number of tunable model parameters, as well as networks trained on the two feature types: low-level variables and high-level variables.

Hyperparameters for the deep neural network architecture and training algorithm were selected to minimize the expected generalization loss using the Spearmint Bayesian optimization algorithm~\cite{snoek_practical_2012, gelbart_bayesian_2014}. The algorithm was allowed to perform 100 experiments; each experiment tested a unique combination of hyperparameters, training a neural network on a training data set of 40 million random examples and computing the generalization error on another 10 million random examples. The following hyperparameters were optimized simultaneously: the learning rate decay factor, the initial momentum, the final momentum, the number of epochs until momentum saturation, the number of hidden layers, and the number  of neurons per hidden layer. All hidden layers were constrained to have the same number of neurons. Additional details regarding the hyperparameter search space can be found in the Methods section.

\begin{figure}
\centering
\includegraphics[width=3in]{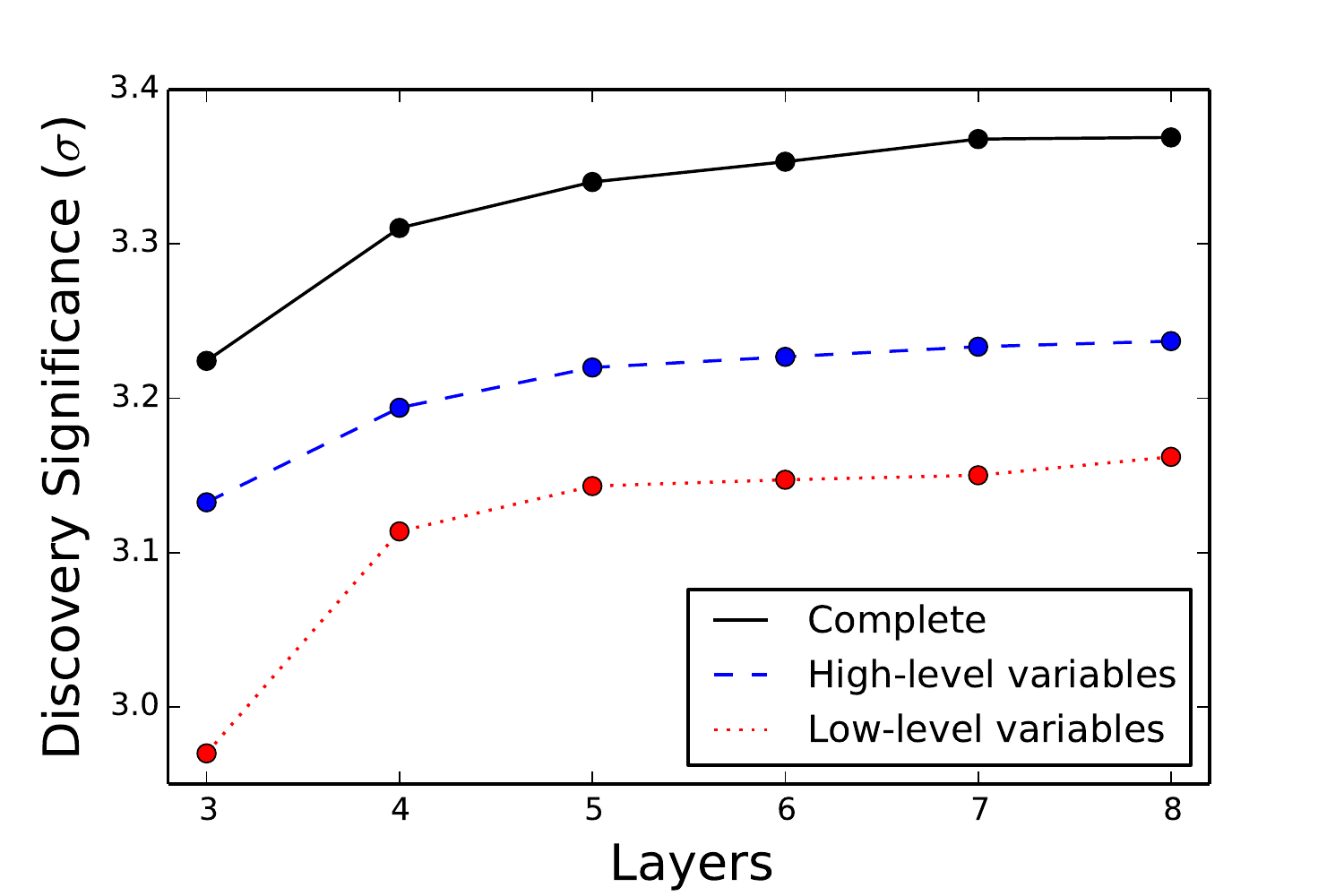}
\caption{Effect of network depth on discovery significance. These networks were trained with the hyperparameters optimized for the deep network.}
\label{fig:ds_vs_nlayers}
\end{figure}

The best hyperparameter combination used the maximum of eight layers. Training was repeated five times using the optimized hyperparameters, random train/test splits, and random weight initializations. Table \ref{tab:auc} shows the mean and variance of two performance metrics; the deep neural networks (DNN) achieve a mean Area Under the Receiver Operator Characteristic Curve (AUC) of 0.802 (s.d. $0.0001$), and the corresponding expected significance of discovery is 3.37 (s.d. $0.003$) Gaussian $\sigma$. The expected significance is calculated using $N_{\rm sig}=100, N_{\rm backg.}=5000\pm 250$ with a profile likelihood method and evaluated with the asymptotic approximation~\cite{Cowan:2010js}.  An additional performance boost is obtained by creating an ensemble classifier from the five networks, which achieves an AUC of 0.803 and discovery significance of $3.39\sigma$.  Figure~\ref{fig:ds_vs_nlayers} shows the effect of network depth on performance.

\begin{figure}
\centering
\includegraphics[width=3in]{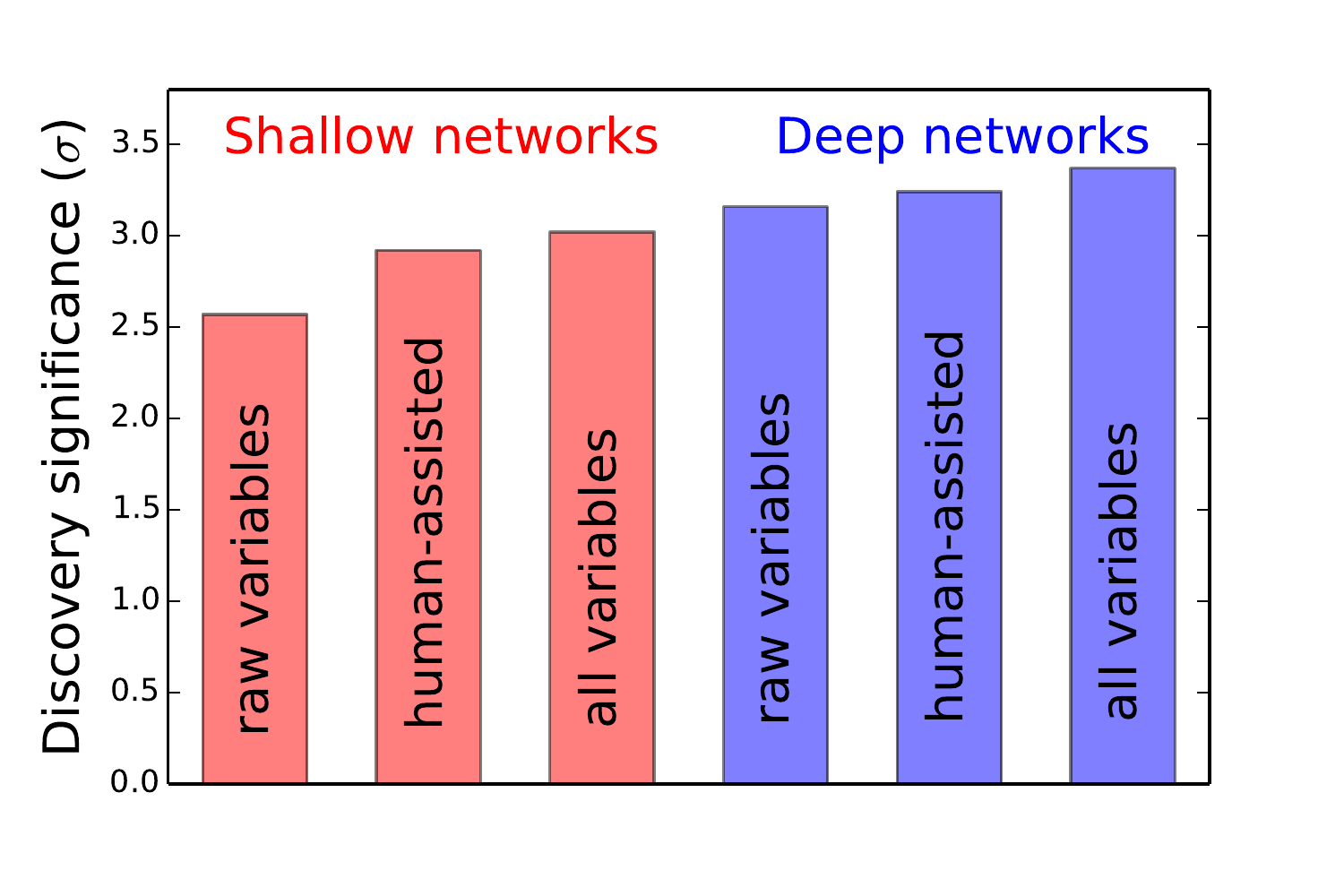}
\caption{ Comparison of discovery significance  for the traditional learning method (left) and the deep learning method (right) using the low-level variables, the  high-level variables and the complete set of variables.}
\label{fig:bar}
\end{figure}

For comparison, we performed the same Bayesian optimization on a set of shallow neural networks (NN). The search space included shallow networks with up to 56,000 hidden units, which have the same number of tunable model parameters as the largest deep networks. However, a single, large, hidden layer did not lead to better performance; the best network had just 691 hidden units. These shallow architectures performed significantly worse than the deep networks, even when using an ensemble (Table \ref{tab:auc}, Figure~\ref{fig:bar}).

The contribution of the high-level variables derived by physicists was analyzed by training on the different feature subsets. Deep and shallow neural networks were trained on both  the 10 low-level variables and  the 15 high-level variables only, using the same hyperparameters that had been optimized for the networks trained on the full feature set. Table \ref{tab:auc} shows that the networks perform better with the high-level variables, but that they perform best with the complete set.  To put into practical context the impact of the boost in discovery significance between the NN and the DNN, we measure the increase needed in the size of the expected data set (nominally $N_{\rm sig}=100, N_{\rm backg.}=5000\pm 250$) to achieve the same enhancement in discovery significance (3.02$\sigma \rightarrow 3.37\sigma$) for the NN; an enlargement of 25\% is required. Therefore, using the DNN dramatically shortens the time needed to operate the collider before the data are statistically significant.

\begin{table}
\centering
\caption{Comparison of the performance of shallow neural networks (NN), and deep neural networks (DNN) for three sets of input features: low-level variables, high-level variables, and the complete set of variables. Each neural network was trained five times with different random weight initializations and different train/test splits. Performance of an ensemble classifier using all five instances is also shown. The table displays the mean Area Under the Curve (AUC) of the signal-rejection curve calculated from 10 million test points; standard deviations are in parentheses. The mean expected significance of a discovery (in units of Gaussian $\sigma$) is given for 100 signal events and 5000 background events with a 5\% relative uncertainty. }
\label{tab:auc}
\begin{tabular}{llll}
\hline\hline
 & \multicolumn{3}{c}{AUC}\\
Technique & Low-level & High-level & Complete \\
\hline
NN 	& $0.789$ $(0.0010)$ & $0.792$ $(0.0002)$ & $0.797$ $(0.0004)$ \\
NN	ensemble & $0.791$ & $0.793$ & $0.798$ \\
DNN	& $0.798$ $(0.0001)$ & $0.798$ $(0.0001)$ & $0.802$ $(0.0001)$ \\
DNN	ensemble & $0.798$ & $0.798$ & $0.803$ \\
\hline\hline
 & \multicolumn{3}{c}{Discovery significance}\\
Technique & Low-level & High-level & Complete \\
\hline
NN 	& $2.57\sigma$ $(0.006)$ & $2.92\sigma$ $(0.006)$ & $3.02\sigma$ $(0.008)$ \\
NN	ensemble & $2.61\sigma$ & $2.96\sigma$ & $3.06\sigma$ \\
DNN  & $3.16\sigma$ $(0.003)$ & $3.24\sigma$ $(0.003)$ & $3.37\sigma$ $(0.003)$ \\
DNN	ensemble & $3.18\sigma$ & $3.26\sigma$ & $3.39\sigma$ \\
\hline\hline
\end{tabular}
\end{table}

\section{Discussion}

As expected, the high-level variables derived by physicists clearly capture features of the data that are useful for classification. But while the shallow neural networks perform poorly on the low-level variables alone, deep networks trained on the low-level variables perform nearly as well as deep networks trained on the complete set of variables (even though the Bayesian hyperparameter optimization was performed on networks trained with the complete set). The deep networks are able to learn most of the discriminative information contained in the high-level variables from the low-level variables alone.  Note that in the case of the high-level $m_{\rm MMC}$ variable, information regarding the known mass of the $\tau$ lepton is included in the calculation; this information is not available from the low-level variables, which may explain in part why the DNN with only low-level variables does not completely match the performance of the DNN with the complete set.

It is also interesting to note that the deep networks learn \emph{additional} information from the low-level variables that is not contained in the high level variables, as evidenced by the deep networks performing better than the shallow networks on both the low-level variables and the complete variables; in fact the deep network with the low-level variables alone performs better than the shallow network trained on the complete set.

A competition hosted through Kaggle recently challenged participants to build classifiers for a similar  machine learning task, where one $\tau$ lepton decays hadronically to a jet of mostly visible particles. However, the Kaggle data set contains slightly different features and only 250 thousand training examples, compared to the 40 million training examples used here. To avoid overfitting with such a small training set, one would have to use very small or heavily-regularized classifiers. Since an arbitrary quantity of training data can be generated through simulations, this is an unnecessary handicap.

\section{Conclusion}
We have demonstrated that deep neural networks lead to significantly better performance on detecting $H\rightarrow\tau^+\tau^-$ decay events from background compared to shallow networks containing the same number of tunable parameters and optimized in the same manner. Furthermore, the Bayesian optimization algorithm decided for itself that the best neural network depth was \emph{eight layers}, the maximum that we had set prior to starting the algorithm. The deep networks trained on the low-level variables performed \emph{better} than shallow networks trained on the high-level variables engineered by physicists, and almost as well as the deep networks trained high-level variables, suggesting that they are \emph{automatically} learning the discriminatory information contained in the physicist-derived high-level variables.   The improvement in discovery significance is equivalent to an increase in the accumulated dataset of 25\%.

\section{Methods}

Neural network classifiers were trained with rectified linear hidden units, a logistic output unit, and cross-entropy loss. Network parameters were trained using stochastic gradient descent with mini-batches of 100 examples. A momentum term increased linearly from an initial value to some final value over a specified number of epochs. The entire dataset of 80 million samples was normalized prior to any training; for those features with a skewness greater than $1.0$, a small value of $10^-8$ was added, the logarithm was taken, then the values were normalized.

Computations were performed using machines with 16 Intel Xeon cores, 64 GB memory, and NVIDIA Titan or Tesla C2070 graphics processors. All neural networks were trained using the Pylearn2 and Theano software packages~\cite{bergstra_theano:_2010,goodfellow_pylearn2:_2013}. Bayesian optimization was performed with the Spearmint software package using a Gaussian Process model, running 20 experiments in parallel, with 20 Markov Chain Monte Carlo iterations and a burn-in of 50. The test cross-entropy error was used as the objective function for the Bayesian optimization.

The space of possible hyperparameters for the deep networks had six dimensions: number of layers (2 to 8), number of hidden units per layer (100 to 500), learning rate decay factor ($1 + 10^{-9}$ to $1 + 10^{-3}$, log scaled), initial momentum (0 to 0.5), final momentum (0.001 to 0.5, log scaled), epochs until momentum saturation (20 to 100). The initial learning rate was $0.01$. The space of possible hyperparameters for the shallow networks also had six dimensions: number of hidden units per layer (100 to 56234, log scaled), initial learning rate ($10^{-7}$ to $10^{-2}$, log scaled), learning rate decay factor ($1 + 10^{-9}$ to  $1 + 10^{-3}$, log scaled), initial momentum (0 to 0.5), final momentum (0.001 to 0.5, log scaled), epochs until momentum saturation (20 to 100). 

The best single deep network used eight layers, with 274 hidden units in each of the seven hidden layers. The momentum parameter began at $0$ and increased linearly before saturating at $0.996$ by epoch $13$. The learning rate decayed by a factor of $1.00000051371$ after each mini-batch update. Experiments with additional hidden layers did not improve performance, but the parameter space for deeper networks was not explored thoroughly. The best shallow network had a single hidden layer of $693$ rectified linear units, with an optimized initial learning rate of $0.006$, a learning rate decay factor of $1.00000000894$, and the momentum remaining constant at $0.5$ throughout training. The initial weights of both the shallow and deep networks were drawn from a normal distribution with zero mean and a standard deviation of $0.1$ for the first layer, $0.001$ for the last layer, and $1.0$ divided by the square root of the number of inputs for the other layers.

Experiments using the dropout algorithm and other stochastic methods like adding Gaussian noise to the neuron activations~\cite{hinton_improving_2012,baldi_dropout_2014} did not improve performance. These approaches help with regularization, but they generally make it more difficult for the network to fit to the training data. Our training data set is large, so the primary challenge is learning   rather than avoiding overfitting.

\section*{Acknowledgments} 
We are grateful to David Rousseau and Kyle Cranmer for their insightful comments, Michael Gelbart for his help with the Bayesian optimization, and Yuzo Kanomata for computing support. We also wish to acknowledge a hardware grant from NVIDIA, NSF grant IIS-0513376, and a Google Faculty Research award to PB.

\section*{References}
\bibliography{paper,2014DeepNN,physics,baldi,nn,atlas}

\end{document}